\documentstyle[preprint,pre,aps,eqsecnum]{revtex}

\newcommand{\real}{{\sf I}\kern-.12em{\sf R}}

\begin{document}

\rightline{IFUP-TH 58/95}
\rightline{UCY-PHY 96/9}

\centerline{\bf The three-loop $\beta$ function in $SU(N)$
lattice gauge theories}
\vskip 5mm
\centerline{B. All\'es$^1$, A. Feo$^{1,2}$ and H. Panagopoulos$^3$}
\centerline{\it $^1$Dipartimento di Fisica dell'Universit\`a and INFN,
Pisa, Italy}
\vskip 2mm
\centerline{\it $^2$Scuola Normale Superiore, Pisa, Italy}
\vskip 2mm
\centerline{\it $^3$Department of Natural Sciences, University of Cyprus,}
\centerline{\it P.O. Box 537, Nicosia CY-1678, Cyprus }

\begin{abstract}
We calculate the third coefficient of the lattice $\beta$ function
in pure Yang-Mills theory. We make use of a computer code
for solving perturbation theory analytically on the lattice.
We compute the divergent integrals by using a method based on
a Taylor expansion of the integrand in powers of the external momenta
in $4 - \epsilon$ dimensions.
Our results are in agreement with a
previous calculation by M. L\"uscher and P. Weisz where the authors
used a different technique.
We also show how this new coefficient modifies the scaling function on
the lattice in both the standard and energy schemes.
In particular we show that asymptotic scaling is extremely
well achieved in the energy scheme.
\end{abstract}

\vskip 5mm


\vfill\eject

\section{Introduction}

\vskip 5mm

The relationship between the bare coupling and the
cutoff is an essential ingredient in lattice calculations.
Usual Monte Carlo simulations require the knowledge of this
relationship far from the critical point, where corrections to
asymptotic scaling could become relevant.

In order to check the relevance of these scaling corrections, in
this paper we present a calculation of the first non-universal
coefficient of the lattice $\beta$ function in pure Yang-Mills theory.
This renormalization group function can be written as
\begin{equation}
\beta^L(g_0) \equiv - a \frac{\hbox{d}g_0}{\hbox{d}a} \mid_{g_r,\mu}=
- b_0^L g_0^3 - b_1^L g_0^5 - b_2^L g_0^7 - \cdots
\end{equation}
where $a$ is the lattice spacing, $g_0$ the lattice bare coupling,
$g_r$ is the renormalized coupling constant and $\mu$ the subtraction
point. It establishes how the bare coupling and the cutoff $a$ must
simultaneously vary to keep the renormalized quantities fixed.
The first two coefficients in eq. (1.1) are known as they are
scheme-independent. Our purpose is to compute the third coefficient,
$b_2^L$.

In a recent paper \cite{lw234} the authors calculated a quantity
related to this coefficient
using a coordinate space method \cite{lw429}
for evaluating the lattice integrals. In our calculation we have used
a different technique where firstly the integrands are Taylor expanded
in powers of the external momenta and then computed term by term with
the introduction of an IR regulator. The idea is a generalization
to two loops of the procedure introduced in \cite{kawai}.
Our result is an independent check of the calculation
in reference \cite{lw234}.
Our integration method can be straightforwardly generalized to the case
with fermions.

In order to facilitate comparison we have adopted the notation of
ref. \cite{lw234}.

The plan of the paper is as follows. In section 2 we will show the
method we followed to determine $b_2^L$. We will give explicit formulae
that relate the coefficient $b_2^L$ to the same coefficient in
the renormalized $\beta$ function (see eq. (2.9)).
In section 3 we give the final result of our calculation as
the two-loop contribution to the coupling
renormalization constant and we derive the value for $b_2^L$
and discuss its consequences for the scaling behaviour of
dimensionful quantities in present-day
Monte Carlo simulations for both the standard and energy schemes.
Finally, in section 4 we introduce our integration method by
solving a typical two-loop integral. The conclusions are presented in
section 5. A list of superficially divergent integrals used in our
computation are shown in the Appendix.

The involved algebra of the lattice perturbation theory was
carried out by making use of a computer code.
The main lines of this code were explained
in \cite{npb} and used
to compute the three-loop perturbative background of the
topological susceptibility in pure Yang-Mills on the lattice and the
three-loop lattice free energy in the same theory, \cite{plb}.
 For the purpose of the present paper, this code was extended to
include form factors.

\vskip 1cm

\section{A two-loop calculation}

\vskip 5mm

Although $b_2^L$ is a three-loop quantity, it can be computed
by evaluating two-loop diagrams on the lattice
provided the corresponding coefficient in the renormalized $\beta$
function, $b_2$, is known.
The renormalized $\beta$ function is defined as
\begin{equation}
\beta(g_r) \equiv \mu \frac{\hbox{d}g_r}{\hbox{d}\mu} \mid_{g_0,a}=
- b_0 g_r^3 - b_1 g_r^5 - b_2 g_r^7 - \cdots
\end{equation}
This function depends on the renormalization scheme only,
hence it must be the same for any regularization.
In this section we will show how the coefficient $b^L_2$ can be
derived from this $\beta$ function.

The bare coupling constant on the lattice
$g_0$ and the renormalized one $g_r$ (for instance in the
$\overline{\rm MS}$
scheme in the continuum) are related by
\begin{equation}
g_0 = Z(g_0,\mu a) g_r \\
\end{equation}
where
\begin{eqnarray}
Z(g_0,\mu a) &=& 1 + Z_1 g_0^2 + Z_2 g_0^4 + \cdots \nonumber \\
Z_1 = Z_{10} + Z_{11} &\ln \mu a& \qquad \qquad
Z_2 = Z_{20} + Z_{21}
\ln \mu a + Z_{22} \ln^2 \mu a
\end{eqnarray}
The $\beta^L$ function (1.1) can be computed from the previous
renormalization constant by calculating the derivative of (2.2)
with respect to the lattice spacing $a$
\begin{equation}
 \beta^L(g_0) = \frac{g_0}{Z(g_0,\mu a)} \left(
\frac{\partial Z(g_0,\mu a)}{\partial g_0} \beta^L(g_0) -
a \frac{\partial Z(g_0,\mu a)}{\partial a}\right).
\end{equation}
This equation yields $b_0^L = Z_{11}$,
$b_1^L = Z_{21} + Z_{10} Z_{11}$; while the
finiteness of the beta function implies $Z_{11}^2 + 2 Z_{22} = 0$.

Now, combining the derivative of eq. (2.2) with respect to $\mu$
\begin{equation}
0 = \beta(g_r) Z(g_0, \mu a) + g_r \mu \frac{\partial
Z(g_0, \mu a)}{\partial \mu}
\end{equation}
with eq. (2.4) we obtain
\begin{equation}
\beta^L(g_0) = {{ \beta(g_r(g_0)) Z(g_0,\mu a)}
\over\displaystyle {
1 -  {{\partial \ln Z(g_0,\mu a)}\over\displaystyle{\partial \ln g_0}}}}.
\end{equation}
This equation relates the perturbative coefficients in
eqs. (2.1) and (1.1). In
particular we obtain
\begin{eqnarray}
b^L_0 &=& b_0, \\
b^L_1 &=& b_1, \\
b^L_2 &=& b_2 - 2 b_1 Z_{10} + b_0 Z_{10}^2 + 2 b_0 Z_{20}.
\end{eqnarray}
Eqs. (2.7) and (2.8) establish the well-known result that the
first two coefficients are independent of the renormalization scheme.
Eq. (2.9) implies that if we know the value of $b_2$ in some
renormalization scheme, we need only to
perform a two-loop calculation in that scheme
on the lattice to obtain $b^L_2$. Throughout this work we will use the
$\overline{\rm MS}$ scheme.
Had we used eq. (2.4) then a
lattice three-loop calculation would have been needed.
The coefficient $Z_{10}$ was computed in reference \cite{kawai}, its
value for the gauge group $SU(N)$ being
\begin{equation}
Z_{10}=
N \left( {1 \over {96 \pi^2}} + {1 \over {16 N^2}}
- {1 \over 32} - {5 \over 72} P_1 - {11 \over 6} P_2 \right).
\end{equation}
In this expression $P_1$ and $P_2$ are numerical constants,
$P_1=0.15493339$, $P_2=0.024013181$.
The coefficients $b_0$ and $b_1$ are
\begin{eqnarray}
b_0 &=& {11 \over 3} {N \over {16 \pi^2}}, \nonumber \\
b_1 &=& {34 \over 3} \left({N \over {16 \pi^2}}\right)^2.
\end{eqnarray}
Moreover, the coefficient $b_2$ in the $\overline{\rm MS}$ scheme is
\cite{tarasov}
\begin{equation}
b_2 = {2857 \over 54} \left({N \over {16 \pi^2}}\right)^3.
\end{equation}

The computation of
the coupling renormalization constant, eq. (2.3), is easier in
the background field gauge.
In fact, this renormalization constant
has a simple relationship with the
background field renormalization constant \cite{abbot},
\begin{equation}
Z^A(g_0,\mu a)   Z(g_0, \mu a)^2 = 1 \qquad \qquad
A_{\mu} = Z^A(g_0,\mu a)^{1/2} A_{r\mu} .
\label{eq:zeqza}
\end{equation}
Relation (2.13) has been recently proven also on the lattice
\cite{lw213}.
Hence we need to compute only a two-loop self-energy of the background
field on the lattice.
In our computation we used the Wilson action in the
background field gauge \cite{ellism}. In this formulation the
links are written as
\begin{equation}
U_{\mu}(x) = U^{q}_{\mu}(x) U^{cl}_{\mu}(x), \qquad \qquad
U^{q}_{\mu}(x) \equiv e^{i g_0 Q_{\mu}(x)}, \qquad
U^{cl}_{\mu}(x) \equiv e^{i a g_0 A_{\mu}(x)},
\end{equation}
where $Q_{\mu}(x)=T^c Q^c_{\mu}(x)$ and
$A_{\mu}(x) = T^c A_{\mu}^c(x)$ are the quantum and
background fields respectively.
The following gauge-fixing term preserves gauge invariance of the
background field
\begin{eqnarray}
S_{gf} &=& \frac{1}{\alpha_0} \sum_{\mu , \nu} \sum_{x}
\hbox{Tr} \, D^-_{\mu} Q_{\mu}(x) D^-_{\nu} Q_{\nu}(x), \\
D^-_{\mu} Q_{\nu}(x) &\equiv & U^{cl-1}_{\mu}(x -
{\hat \mu}) Q_{\nu}(x - {\hat \mu}) U^{cl}_{\mu}(x - {\hat \mu}) -
Q_{\nu}(x).
\end{eqnarray}
We worked in the Feynman gauge, $\alpha_r = 1$.
This gauge fixing produces the following
Fadeev Popov action for the ghosts fields
$\omega$ and $\overline\omega$
\begin{eqnarray}
S_{gh} &=& 2 \sum_{x} \sum_{\mu} \hbox{Tr} \,
(D^+_{\mu}\omega(x))^{\dagger} \Bigl( D^+_{\mu}\omega(x) +
i g_0 \left[Q_{\mu}(x),
\omega(x)\right] + \frac{1}{2}
i g_0 \left[Q_{\mu}(x), D^+_{\mu}\omega(x) \right] \nonumber\\
 & & \quad - \frac{1}{12}
g_0^2 \left[Q_{\mu}(x), \left[ Q_{\mu}(x),
D^+_{\mu}\omega(x)\right]\right] + \cdots \Bigr), \\
D^+_{\mu}\omega(x) &\equiv & U^{cl}_{\mu}(x) \omega(x + {\hat \mu})
U^{cl-1}_{\mu}(x) - \omega(x).
\end{eqnarray}
Finally the change of integration variables from links to vector
fields yields a jacobian that can be rewritten as the measure action
\begin{equation}
S_{m} = \frac{1}{12} N g_0^2 \sum_{x} \sum_{\mu} \hbox{Tr} \,
Q_{\mu}(x) Q_{\mu}(x) + \cdots
\end{equation}
In all expansions, eqs. (2.17) and (2.19), we have written only
the relevant terms for our two-loop computation.

The full action is therefore $S = S_{Wilson} + S_{gf} + S_{gh} + S_m$.
The vertices needed in perturbation theory are obtained as usual
by expanding the exponential of the previous action in powers of
the fields. In ref. \cite{npb} we described in some detail our automatic
procedure for generating vertices.

The Feynman diagrams contributing to the background gluon self-energy
are shown in Figure 1. The bare and renormalized
self-energies of the background field are related by eq.
(\ref{eq:zeqza})
\begin{equation}
(1 - \nu_r(p,\mu,g_r) ) = Z^A (1 - \nu(p,a,g_0) ).
\label{eq:selfs}
\end{equation}
In the computation we used the bare quantum fields. Therefore the
gauge parameter must be explicitly renormalized. Diagrams $1-4$
correspond to this renormalization.
Up to two loops the bare self-energy can be written as
\begin{equation}
\nu(p,a,g_0) = g_0^2 \nu^{(1)} + g_0^4 \nu^{(2)} + \alpha_r
\left( Z_{10}^Q + Z_{11}^Q \ln \mu a \right)
{{\partial \nu^{(1)}} \over {\partial \alpha_0}} \vert_{\alpha_0
= \alpha_r},
\label{eq:pipi}
\end{equation}
where the relation $\alpha_0=Z^Q \alpha_r$ has been used. $Z^Q$ is the
quantum field renormalization constant. We need this constant at
one-loop, $Z^Q = 1 + g_0^2 ( Z^Q_{10} +  Z^Q_{11} \ln \mu a)$
\cite{kawai},
\begin{eqnarray}
Z^{Q}_{11} &=& - {10 \over 3}
{N \over {16 \pi^2}}, \nonumber \\
Z^{Q}_{10} &=& N \left(
{7 \over 72} P_1 + {1 \over 16}
- {1 \over {8 N^2}} - {1 \over {48 \pi^2}}
+ {5 \over 3} P_2 \right).
\end{eqnarray}
The last term in eq. (\ref{eq:pipi}) corresponds to the diagrams $1-4$
in Figure 1.

The renormalized self-energy can be written similarly as
$\nu_r(p,\mu,g_r) = g_r^2 \nu_r^{(1)} + g_r^4 \nu_r^{(2)}~+~\cdots$.
By using this relation and eqs.
(2.3), (\ref{eq:zeqza}), (\ref{eq:pipi}) we obtain
\begin{equation}
2 Z_2 + (Z_1)^2 = \nu_r^{(2)} - \nu^{(2)} -
\left( Z_{10}^Q + Z_{11}^Q \ln \mu a \right) \alpha_r
{{\partial \nu_r^{(1)}} \over {\partial \alpha_r}} .
\end{equation}
In this expression, we took advantage of the gauge parameter
independence of the coupling renormalization constant $Z$ in the
$\overline{\rm MS}$ scheme to substitute $\nu^{(1)}$ by $\nu_r^{(1)}$
in the derivative~\cite{caswell}.

The explicit expressions for the one-loop and two-loop renormalized
self-energy of the background field were computed in ref \cite{ellis}.
In the $\overline{\rm MS}$ scheme they are
\begin{eqnarray}
\nu_r^{(1)} &=& {N \over {16 \pi^2}} \left(
 - {11 \over 3} \ln {p^2 \over \mu^2} + {205 \over 36} + {3 \over 2}
\alpha_r + {1 \over 4} \alpha_r^2 \right) \\
\nu_r^{(2)} &=& \left( {N \over {16 \pi^2}} \right)^2
\left( - 8 \ln {p^2 \over \mu^2} + {577 \over 18} - 6 \zeta (3)
 \right).
\end{eqnarray}
Here $\zeta (3)=1.202057$ is the Riemann's zeta function and $\mu$ is
the renormalization point. The two-loop
self-energy (2.25) is written in the Feynman gauge.

\vskip 1cm

\section{The results}

\vskip 5mm

The result for each diagram can be written as $A_{\rho}(p) A_{\sigma}(-p)
{\sf T}_{\rho\sigma}(p) $ with ${\sf T}_{\rho\sigma}(p)$ some
tensor. Leaving out the fields and summing over $\rho$, $\sigma$ we obtain
${\rm Tr}{\sf T}(p)$.
The gauge invariance of the final result guarantees that the sum
of all diagrams yields $3 p^2 \left(1 -\nu(p,a,g_0) \right)$.

For some diagrams we have computed the whole expression $A_{\rho}(p) A_{\sigma}(-p)
{\sf T}_{\rho\sigma}(p) $ for purposes of
checking. In all cases we obtain agreement (within the precission used) with
\cite{lw234}. Therefore we do not show the results for each diagram.

After collecting all individual contributions
we get for $\nu^{(2)}$ the following expression
\begin{eqnarray}
\nu^{(2)} &=& -{N^2 \over {32 \pi^4}} \ln (a^2 p^2) +
{3 \over {128\,N^2}} +c_1 + c_2 \, N^2 \nonumber \\
c_1 &=& - 0.01654 \nonumber \\
c_2 &=& 0.007230.
\end{eqnarray}
Now, using eq. (2.23) we get the two-loop term of the coupling
renormalization constant $Z(g_0, \mu a)$ in eq. (2.3)
\begin{eqnarray}
Z_2 &=& \left( {1 \over {16 \pi^2}} \right)^2
\Bigg( 358.5 - {{340.9} \over N^2} - 188.7 \,N^2 + \nonumber \\
&& \left( 
N^2 \left( {193 \over 18} + {11 \over 6} \pi^2 +
{110 \over 27} \pi^2 P_1 + {968 \over 9} \pi^2 P_2 \right)
- {11 \over 3} \pi^2 \right) \ln \mu a -
{{121 \, N^2} \over {18}} \ln^2 \mu a \Bigg).
\label{eq:zfinal}
\end{eqnarray}
Notice that as predicted $Z_{11}^2 + 2 Z_{22}$
vanishes. The coefficient of the
logarithm in the previous equation contributes to
the second coefficient of
the $\beta$ function, $b^L_1$. The other term in eq. (\ref{eq:zfinal})
together with eqs. (2.9) and (2.10-2.12)
gives the desired result for $b^L_2$,
\begin{eqnarray}
b_2^L &=& \left( {N \over {16 \pi^2}} \right)^3
\Bigg( -{{22\,\pi^4} \over N^4} - {1 \over N^2} \,\left(
8 \pi^2 + {{2816} \over 3}\pi^4\,c_1 \right) + \nonumber \\
&& {{1522} \over 9} + 4\,\pi^2  - {{2816} \over 3}\pi^4\, c_2 +
{{124} \over 9}\pi^2\,P_1 + {{1408}\over 3}\pi^2 \,P_2 -
22 \, \zeta (3) \Bigg)
\end{eqnarray}
and numerically
\begin{equation}
b_2^L = \left( {N \over {16 \pi^2}} \right)^3
\left( -366.2 + {{1433.8} \over {N^2}} - {{2143.} \over {N^4}} \right).
\end{equation}
Now, by integrating the lattice
$\beta$ function, eq. (1.1), we get the
following dependence of the lattice cutoff on the bare coupling
\begin{equation}
a \Lambda_L = \exp\left(-1/2 b_0 g_0^2 \right)
(b_0 g_0^2)^{-b_1/2 b_0^2} \left(
1 + {1 \over {2 b_0^3}} \left( b_1^2 - b_2^L b_0 \right) g_0^2
+ \cdots \right),
\label{eq:betaint}
\end{equation}
where $\Lambda_L$ is the renormalization group invariant mass in the
lattice regularization. For the gauge group $SU(3)$
eq. (\ref{eq:betaint}) becomes
\begin{equation}
a \Lambda_L = \left({{16 \pi^2} \over {11 g_0^2}}\right)^{51/121}
\exp \left( - {{8 \pi^2} \over {11 g_0^2}} \right) \left( 1 +
0.1896 g_0^2 \right).
\label{eq:lusresult}
\end{equation}
Thus in the region $g_0 \approx 1$ the last factor in eq.
(\ref{eq:lusresult}) brings about a $\sim 20\%$ correction.
In the $SU(2)$ gauge group this correction amounts to $0.08324\,g_0^2$.
At large $N$ the correction behaves linearly as
$0.1048 \,N \,g_0^2$.

We can also calculate the correction to asymptotic scaling in the
scheme of the energy~\cite{parisi}. The weak expansion of the
average plaquette is~\cite{plb,gcrossi}
\begin{eqnarray}
&\langle & 1 - {1\over N} \Box \,\rangle = w_1 g_0^2 +
w_2 g_0^4 + w_3 g_0^6 + \cdot \cdot \cdot \nonumber \\
&w_1& = {{N^2 - 1} \over {8 \,N}} \qquad \qquad
w_2 = {{N^2 - 1} \over {4}} \left(
0.02043 - {1 \over {32\, N^2}} \right) \nonumber \\
&w_3& = {{N\,\left( N^2 - 1\right)} \over {8}}
\left( 0.006354 - {{0.01812} \over N^2} + {{0.01852} \over N^4}
\right).
\label{eq:weak}
\end{eqnarray}
The coupling constant in this scheme can be defined as
\begin{equation}
g_E^2 \equiv {1 \over w_1}  \langle  1 - {1\over N} \Box \,\rangle
= g_0^2 + {w_2 \over w_1} g_0^4 + {w_3 \over w_1} g_0^6 + \cdot \cdot \cdot
\label{eq:ge}
\end{equation}
With this definition, the third coefficient of the beta function is
\begin{equation}
b^E_2 = b_2^L + {{b_0 \,w_3 - b_1\, w_2 - b_0 \, w_2^2/w_1} \over w_1},
\end{equation}
and the scaling function in $SU(3)$ becomes
\begin{eqnarray}
a \Lambda_E &=& \left({{16 \pi^2} \over {11 g_E^2}}\right)^{51/121}
\exp \left( - {{8 \pi^2} \over {11 g_E^2}} \right) \left( 1 +
0.01161 g_E^2 \right) \nonumber \\
\Lambda_E &=& \exp\left({w_2 \over {2 b_0 w_1}} \right)\,\Lambda_L =
2.0758 \,\Lambda_L.
\label{eq:scalingenergy}
\end{eqnarray}
In $SU(2)$ the correction is $0.003530\, g_E^2$.

The previous results show that as expected the first correction to asymptotic
scaling in the scheme of the energy is much smaller than in the usual standard
scheme. This fact indicates that in the scheme
defined by eq. (\ref{eq:ge}) asymptotic scaling is possibly
satisfied within a few per cent at bare couplings as large as $g_E \approx 1$.

\vskip 1cm

\section{The integration method}

\vskip 5mm

In this section we introduce our integration method.
First we show how we computed UV-divergent integrals and then we will
say a few words about the convergent integrals.

The method for UV-divergent integrals will be shown by solving the
following example
\begin{equation}
E(ap) \equiv \int^{+\pi}_{-\pi} {{\hbox{d}^4 k} \over {(2 \pi)^4}}
\int^{+\pi}_{-\pi} {{\hbox{d}^4 q} \over {(2 \pi)^4}}
{1 \over {{\hat {\mathstrut q}}^2 {\hat {\mathstrut k}}^2
{\widehat {\mathstrut {q + k}}}^2
{\widehat {\mathstrut {k + ap}}}^2}}.
\end{equation}
Here $p$ is the
external momentum and $\hat q^2 \equiv 4 \sum_{\mu}
\sin^2(q_{\mu}/2)$.
This integral diverges when $a \rightarrow 0$. Similar one-loop
divergent integrals can be computed by the following technique
\cite{kawai} (see also ref. \cite{vicari}): {\it i)} define
the same integral in $D>4$ dimensions; {\it ii)} Taylor expand
this integral in the external momenta until the remainder of the
expansion is UV-finite; {\it iii)} integrate all terms in this
expansion including the remainder and taking
into account the IR-divergences that are generated.
The remainder can be computed in
the continuum as it is UV-finite. Eventually all IR-poles must cancel
giving an IR-finite expression.
Notice that the computation of the remainder in the continuum requires
an exchange of the limits $a \rightarrow
0$ and $\epsilon_{IR} \rightarrow 0$; this delicate operation is valid
provided adequate subtractions are made on the integrand.

The method we used to compute the integral (4.1) is a generalization
of this idea beyond one loop. We must simply apply the same
steps to {\it every} loop. Then the integral in eq. (4.1) becomes
($\kappa$ is an arbitrary mass scale)
\begin{eqnarray}
E(ap) &=& (\kappa a)^{4 \epsilon}
\int^{+\pi}_{-\pi}
{{\hbox{d}^D k} \over {(2 \pi)^D}}
\int^{+\pi}_{-\pi} {{\hbox{d}^D q} \over {(2 \pi)^D}}
\left( {1 \over {{\hat {\mathstrut q}}^2 {\widehat {\mathstrut {q+k}}}^2}} -
{1 \over { \left({\hat q}^2\right)^2}} \right)
\left( {1 \over {{\hat {\mathstrut k}}^2
{\widehat {\mathstrut {k + ap}}}^2}} -
{1 \over { \left({\hat k}^2\right)^2}} \right) + \\
& & (\kappa a)^{4 \epsilon}
\int^{+\pi}_{-\pi} {{\hbox{d}^D k} \over {(2 \pi)^D}}
\int^{+\pi}_{-\pi} {{\hbox{d}^D q} \over {(2 \pi)^D}}
\left( {1 \over { {\hat {\mathstrut k}}^2
{\widehat {\mathstrut {k + ap}}}^2{ \left({\hat q}^2\right)^2}}} +
{1 \over { {\hat {\mathstrut q}}^2
{\widehat {\mathstrut {q + k}}}^2{ \left({\hat k}^2\right)^2}}} -
{1 \over { { \left({\hat k}^2\right)^2}  { \left({\hat q}^2\right)^2}}}
\right),
\end{eqnarray}
where we put $D=4 - 2 \epsilon$ (from now on we will omit the subscript
$IR$ to the $\epsilon$). Now the integral in (4.2) is UV-finite and we
can exchange the UV and IR limits to rewrite it as
\begin{equation}
\kappa^{4 \epsilon} \int_{{\real}^D}
{{\hbox{d}^D k} \over {(2 \pi)^D}}
\int_{{\real}^D} {{\hbox{d}^D q} \over {(2 \pi)^D}}
{{p (p + 2 k)} \over {(k + p)^2 (k^2)^2}}
{{k (k + 2 q)} \over {(k + q)^2 (q^2)^2}} + {\cal O}(a),
\label{eq:p1}
\end{equation}
which can be computed by continuum methods.

The first and third terms in eq. (4.3) can be combined to give
\begin{equation}
(\kappa a)^{4 \epsilon}
\int^{+\pi}_{-\pi} {{\hbox{d}^D q} \over {(2 \pi)^D}}
{1 \over { ({\hat q}^2)^2}} \int_{-\pi}^{+\pi}
{{\hbox{d}^D k} \over {(2 \pi)^D}}
\left( {1 \over { {\hat k}^2 {\widehat {k + ap}}^2}} -
{1 \over { ({\hat k}^2)^2}} \right).
\label{eq:p2}
\end{equation}
The $k$-integration is UV-finite, therefore we can solve this
integral by perfoming the corresponding continuum limit.

The second term in eq. (4.3) is IR-divergent and can be calculated by
performing a chain of adequate additions and subtractions.
We can rewrite it as
\begin{eqnarray}
&(\kappa a)^{4 \epsilon} &
\int^{+\pi}_{-\pi} {{\hbox{d}^D k} \over {(2 \pi)^D}}
{1 \over { ({\hat k}^2)^2}} \int_{-\pi}^{+\pi}
{{\hbox{d}^D q} \over {(2 \pi)^D}}
\left( {1 \over { {\hat q}^2 {\widehat {q + k}}^2}} -
{1 \over { ({\hat q}^2)^2}} \right) + \label{eq:p3prev} \\
& (\kappa a)^{4 \epsilon} &
\int^{+\pi}_{-\pi} {{\hbox{d}^D k} \over {(2 \pi)^D}}
\int^{+\pi}_{-\pi} {{\hbox{d}^D q} \over {(2 \pi)^D}}
{1 \over { ({\hat q}^2)^2}} {1 \over { ({\hat k}^2)^2}}.
\label{eq:p3}
\end{eqnarray}
This operation has not modified the bad IR and UV properties of the
integrand. However, eq. (\ref{eq:p3})
can be computed numerically as we
will show below. Let us study eq. (\ref{eq:p3prev}).
We would know how to solve this integration if it was written in the
continuum. Following this hint we add and subtract the
analogous expression in the continuum for the integration in $q$ to
get
\begin{eqnarray}
&(\kappa a)^{4 \epsilon} &
\int^{+\pi}_{-\pi} {{\hbox{d}^D k} \over {(2 \pi)^D}}
{1 \over { ({\hat k}^2)^2}} \int_{-\pi}^{+\pi}
{{\hbox{d}^D q} \over {(2 \pi)^D}}
\left( {1 \over { {\hat q}^2 {\widehat {q + k}}^2}} -
{1 \over { ({\hat q}^2)^2}} - {1 \over {q^2 (q+k)^2}} +
{1 \over {(q^2)^2}} \right) + \label{eq:p4} \\
&(\kappa a)^{4 \epsilon} &
\int^{+\pi}_{-\pi} {{\hbox{d}^D k} \over {(2 \pi)^D}}
{1 \over { ({\hat k}^2)^2}} \int_{-\pi}^{+\pi}
{{\hbox{d}^D q} \over {(2 \pi)^D}}
\left( {1 \over {q^2 (q+k)^2}} - {1 \over {(q^2)^2}} \right).
\label{eq:p4post}
\end{eqnarray}
As expected, eq. (\ref{eq:p4}) is IR-finite so that it can be computed
numerically in 4 dimensions. Furthermore, the subtraction of fractions
in the $q$-integration of eq. (\ref{eq:p4post})
makes its $k$-integration
IR-finite. Therefore all IR-divergences in eq.
(\ref{eq:p4post}) are due to the
$q^2 \approx 0$ region in the $q$-integration. Hence a possible way to
solve this integral is to split the $q$-integration into two regions
\begin{eqnarray}
&(\kappa a)^{4 \epsilon} &
\int^{+\pi}_{-\pi} {{\hbox{d}^D k} \over {(2 \pi)^D}}
{1 \over { ({\hat k}^2)^2}} \int_{{\real}^D}
{{\hbox{d}^D q} \over {(2 \pi)^D}}
\left( {1 \over {q^2 (q+k)^2}} - {1 \over {(q^2)^2}} \right) -
\label{eq:p5} \\
&(\kappa a)^{4 \epsilon} &
\int^{+\pi}_{-\pi} {{\hbox{d}^D k} \over {(2 \pi)^D}}
{1 \over { ({\hat k}^2)^2}} \int_{{\real}^D - \pi^D}
{{\hbox{d}^D q} \over {(2 \pi)^D}}
\left( {1 \over {q^2 (q+k)^2}} - {1 \over {(q^2)^2}} \right),
\label{eq:p6}
\end{eqnarray}
where ${\real}^D - \pi^D$ means that the integration domain covers
all the euclidean space-time
except for a hypercube of side $\pi$ around
the origin.
Now, eq. (\ref{eq:p6})
is IR-finite and can be evaluated numerically in $D=4$
dimensions. Integral (\ref{eq:p5})
can be solved by continuum techniques.

Some one-loop IR-divergent integrals are needed to ${\cal O}(\epsilon)$. They are
\begin{eqnarray}
\int_{-\pi}^{+\pi} {{\hbox{d}^D q} \over {(2 \pi)^D}} \,
{1 \over { ({\hat q}^2)^2}} &=&
{1 \over {16 \pi^2}} \left(-{1 \over \epsilon} + 1.838199 +
2.283043 \,\epsilon + {\cal O}(\epsilon^2) \right), \nonumber \\
\int_{-\pi}^{+\pi} {{\hbox{d}^D q} \over {(2 \pi)^D}} \,
{1 \over { ({\hat q}^2)^2}} \,(q^2)^{-\epsilon} &=&
{1 \over {16 \pi^2}} \left(-{1 \over {2 \epsilon}} +
3.315103 - 3.6121 \,\epsilon + {\cal O}(\epsilon^2) \right).
\end{eqnarray}
The first one can be evaluated by using Schwinger parameters.
The second one needs first an expansion in powers of
$\epsilon$, then the Schwinger parametrization can be used.

After evaluating the integrals in eqs.
(\ref{eq:p1},\ref{eq:p2},\ref{eq:p3},\ref{eq:p4},\ref{eq:p5},\ref{eq:p6})
and adding up the partial results
we obtain
\begin{equation}
E(ap) = \left( {1 \over {16 \pi^2}} \right)^2
\left( 28.01 - 6.7920 \,\ln \left( a^2 \,p^2 \right) + {1 \over 2}
\ln^2\left( a^2 \,p^2\right) \right).
\end{equation}
As expected, this result is independent of the IR regularization we
used.

In table I we show a list of the main two-loop divergent integrals
used in this work. All entries in this table have
been computed by using the above technique and the results agree with ref.
\cite{lw234}.

All the other superficially divergent integrals appearing in our computation can
be rewritten in terms of this basic set plus some finite integrals.
The finite lattice integrals have been solved either by using a method based on
Schwinger parameters explained in ref. \cite{npb,amsterdam} or by
an extrapolation to infinite lattices from their values on finite lattices.
The extrapolating function was chosen to be
$r_0 + \sum_{n,m} r_{nm} (\ln^{m} L)/L^{n}$ where $L$ is
the lattice size and $n$ and $m$ free parameters.
The sum on $n$ and $m$ runs over a few integer values (typically
$m=0-1$ and $n=2$). The infinite lattice
value of the integral is $r_0$. Both methods agree. We did not push the precision
to more than four or five digits (for practical purposes in numerical simulations,
knowledge of four digits is sufficient).

In the appendix and in table II we give the list of supercifially divergent integrals
encountered during our computation.

\vskip 1cm

\section{Conclusions}

We have calculated the third coefficient of the lattice $\beta$
function for the Yang-Mills theory with gauge group $SU(N)$.
A computer code was used to simplify the algebra of
perturbation theory on the lattice.
We have also introduced a method to compute the two-loop lattice
integrals. The method can be generalized to any
number of loops and with the inclusion of fermions.
Our result agrees with a similar computation of reference
\cite{lw234}.

The corresponding corrections induced by this
new coefficient cannot be neglected
near the scaling window for the standard scheme. For instance the
modification in the $SU(3)$ gauge group
amounts to $\sim 20\%$ when $g_0 \approx 1$.
Instead, in the energy scheme defined by eq. (\ref{eq:ge}) the correction
to asymptotic scaling in $SU(3)$
is of the order of $\sim 1\%$ at $g_E \approx 1$.
The improvement in $SU(2)$ is also significant: a $\sim 8\%$ correction to
asymptotic scaling in the standard scheme versus $\sim 0.3 \%$ in the energy
scheme when $g_E \approx 1$.
This result emphasizes the role of the energy schemes in numerical
simulations which require good asymptotic scaling.


\section{Appendix}

We present in table II a set of superficially divergent two-loop
integrals encountered in our computation. They can be
written in terms of the basic set in table I plus finite integrals by
using algebraic manipulations based on the identity
\begin{equation}
\widehat{r+s}^2 = \hat{r}^2\,+\,\hat{s}^2 \,+2 \, \sum_{\mu} \sin (r_\mu) \sin (s_\mu) \,-\,
8\, \sum_\mu \sin^2 \left({{r_\mu} \over 2}\right) \,\sin^2 \left({{s_\mu} \over 2}\right),
\end{equation}
valid for any internal or external momenta $r,s$.

This table together with table I is a basis for all superficially divergent
integrals with gluonic propagators, depending on one external momentum.
We evaluate these integrals in the continuum limit.
Integrals with more powers of momenta can be evaluated either by
Taylor expansion or, in the presence of subdivergences, by
subtractions as in section 4, leading to a Taylor expandable part
plus a product of one-loop integrals.

Some trivial cases have been omitted (cases with an
inverse propagator in the numerator, and cases containing $\hat k_\mu
C_{\mu\nu}(k) = 0$, see \cite{lw234}).

The notation is as in table I: $q,\,k$ are internal momenta and $p$ is external.
On the other hand, $(r\cdot s)\equiv \sum_\mu \sin r_\mu \sin s_\mu$ and

\begin{eqnarray}
F_{rs}(ap) \equiv
\int^{+\pi}_{-\pi} {{{\rm d}^4k} \over
{(2\pi)^4}} \,\int^{+\pi}_{-\pi}{{{\rm d}^4q} \over {(2\pi)^4}} \,
{{ (r\cdot s)} \over {\hat q^2 \widehat{k{+}q}^2\hat k^2
\widehat{k{+}p}^2}} \nonumber \\
E_{rsuv}(ap) \equiv \int^{+\pi}_{-\pi} {{{\rm d}^4k} \over
{(2\pi)^4}} \,\int^{+\pi}_{-\pi}{{{\rm d}^4q} \over {(2\pi)^4}} \,
{{(r\cdot s)\, (u\cdot v)} \over {\hat q^2 \widehat{k{+}q}^2
(\hat k^2)^2 \widehat{k{+}p}^2}} \nonumber \\
G_{rsuv}(ap) \equiv \int^{+\pi}_{-\pi} {{{\rm d}^4k} \over
{(2\pi)^4}} \,\int^{+\pi}_{-\pi}{{{\rm d}^4q} \over {(2\pi)^4}} \,
{{ (r\cdot s) \,(u\cdot v)} \over { \hat q^2 \widehat{k{+}q}^2
\hat k^2 \widehat{k{+}p}^2 \widehat{q{-}p}^2}}.
\end{eqnarray}


\vskip 5mm

\vskip 1cm

\section{Acknowledgements}

\vskip 5mm

We thank Adriano Di Giacomo and Ettore Vicari
for useful
discussions. B.A. also acknowledges financial support from an
italian INFN grant and A.F. from an italian CNR grant.
H.P. would like to thank the
Theory Group at Pisa for the warm hospitality.


\newpage

\noindent{\bf Figure captions}

\begin{enumerate}

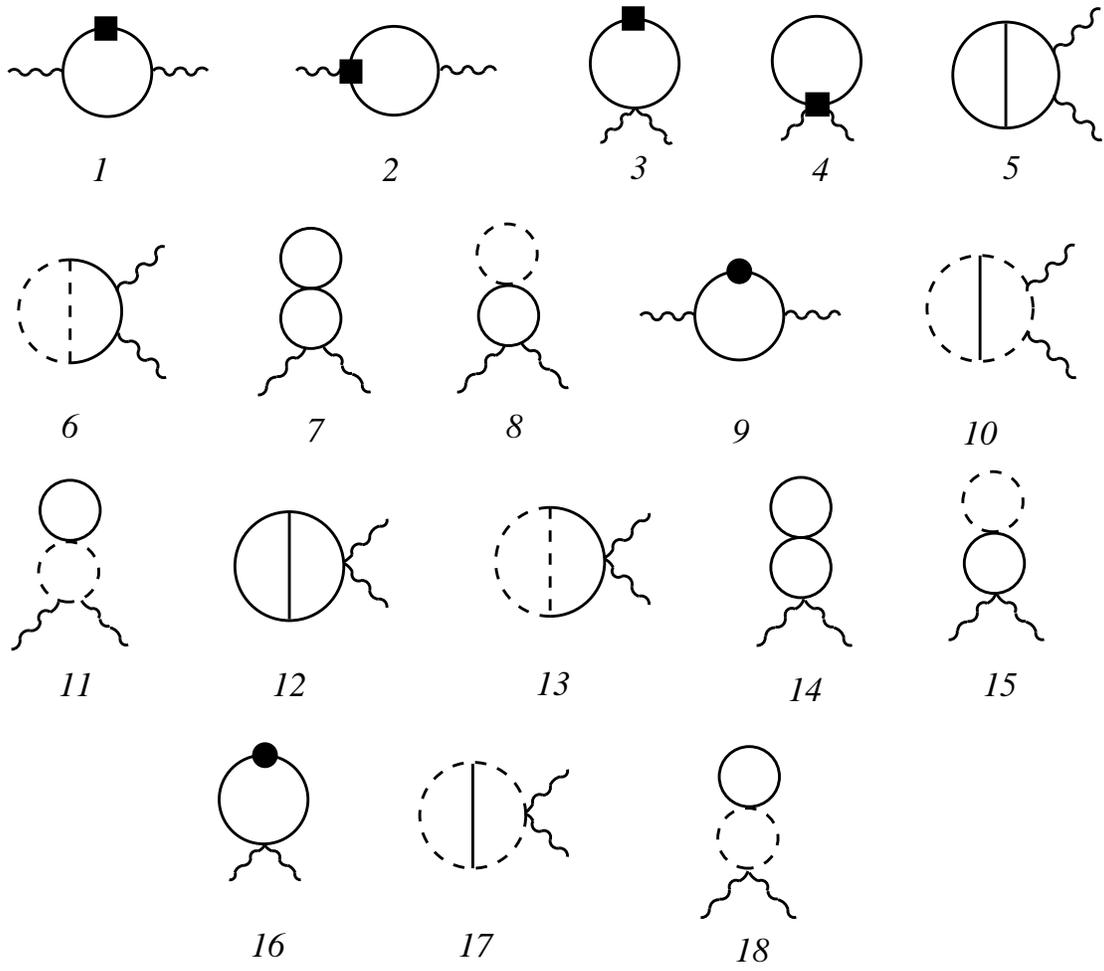
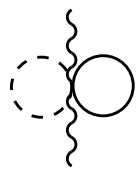
\item[Figure 1.] Set of Feynman diagrams contributing to the two-loop
self-energy of the background field gauge. Straight, dashed and wavy
lines correspond to quantum fields, ghosts and background external
fields respectively. The black circle stands for the measure
vertex and the black square for the counterterms.

\end{enumerate}

\vskip 2cm

\noindent{\bf Table captions}

\vskip 5mm

\begin{enumerate}

\item[Table I.] Basic two-loop divergent lattice integrals.
All expressions in the second column are integrated with the measure
$\int^{+\pi}_{-\pi} \hbox{d}^4q/(2\pi)^4
 \int^{+\pi}_{-\pi} \hbox{d}^4k/(2\pi)^4$. The constant $\zeta (3)
\equiv 1.202057$ is the Riemann's zeta function.

\item[Table II.] Superficially divergent integrals. The notation follows
the appendix. $P_1$ and $P_2$ have been defined in the text.

\end{enumerate}

\newpage

\centerline{\bf Table I}
\vskip 1cm

\moveleft 0.2 in
\vbox{\offinterlineskip
\halign{\strut
\vrule \hfil\quad $#$ \hfil \quad &
\vrule \hfil\quad $#$ \hfil \quad \vrule \cr
\noalign{\hrule}
\hbox{integrand} &
\hbox{result} \cr
\noalign{\hrule}
{1 \over {{\hat {\mathstrut q}}^2 {\hat {\mathstrut k}}^2
{\widehat {\mathstrut {q + k + ap}}}^2 }}
&
\left( {1 \over {16 \pi^2}} \right)^2
\left( 100.8 + a^2\,p^2 \left( -1.857 + {1 \over 2} \ln (a^2\,p^2)
\right) \right)
\cr
\noalign{\hrule}
{1 \over {{\hat {\mathstrut q}}^2 {\hat {\mathstrut k}}^2
{\widehat {\mathstrut {q + k}}}^2
{\widehat {\mathstrut {k + ap}}}^2}}
&
\left( {1 \over {16 \pi^2}} \right)^2
\left( 28.01 - 6.7920 \,\ln (a^2\,p^2) + {1 \over 2} \ln^2 (a^2\,p^2)
\right)
\cr
\noalign{\hrule}
{1 \over {{\hat {\mathstrut q}}^2 {\hat {\mathstrut k}}^2
{\widehat {\mathstrut {q + k}}}^2
{\widehat {\mathstrut {q - ap}}}^2
{\widehat {\mathstrut {k + ap}}}^2}}
&
\left( {1 \over {16 \pi^2}} \right)^2
{1 \over {a^2\,p^2}} \; \; 6 \; \; \zeta (3)
\cr
\noalign{\hrule}
{{\sin k_{\mu} \sin k_{\nu}} \over
{{\hat {\mathstrut q}}^2
{\mathstrut\left( {\hat {\mathstrut k}}^2 \right)^2}
{\widehat {\mathstrut {q + k}}}^2
{\widehat {\mathstrut {k + ap}}}^2}}
&
\left( {1 \over {16 \pi^2}} \right)^2
\Big( {{p_{\mu} p_{\nu}} \over p^2} \left( 2.646 - {1 \over 2}
\ln (a^2\,p^2) \right) +
\cr
&
{1 \over 2} \delta_{\mu\nu} \left(
7.274 - 3.146 \,\ln (a^2\,p^2) + {1 \over 4} \ln^2 (a^2\,p^2) \right)
\Big)
\cr
\noalign{\hrule}
{{\sin q_{\mu} \sin q_{\nu} \cos (k_{\mu}/2) \cos (k_{\nu}/2) } \over
 {{\mathstrut\left( {\hat {\mathstrut k}}^2 \right)^2}
  {\widehat {\mathstrut {k + ap}}}^2 }} \times
&
\left( {1 \over {16 \pi^2}} \right)^2
\Big( {{p_{\mu} p_{\nu}} \over p^2} \left( 0.07838 - {1 \over {24}}
\ln (a^2\,p^2) \right) +
\cr
\left(  ({\widehat {\mathstrut {q + k}}}^2
 {\widehat {\mathstrut {q}}}^2 )^{-1} - 
{\mathstrut\left( {\hat {\mathstrut q}}^2 \right)^{-2}} \right)
&
{1 \over 2} \delta_{\mu\nu} \left(
-1.259 + 0.4435 \,\ln (a^2\,p^2)  - {1 \over {16}} \ln^2 (a^2\,p^2)
\right)
\Big)
\cr
\noalign{\hrule}
}}

\centerline{\bf Table II}
\vskip 1cm

\moveleft 0.2 in
\vbox{\offinterlineskip
\halign{\strut
\vrule \hfil\quad $#$ \hfil \quad &
\vrule \hfil\quad $#$ \hfil \quad \vrule \cr
\noalign{\hrule}
\hbox{integral} &
\hbox{result} \cr
\noalign{\hrule}
F_{pk}(ap)
&
\left( {1 \over {16 \pi^2}} \right)^2 \left( {a^2}\,{p^2}\,\left( -6.774 +
    \left( {5 \over 4} + 8 \pi^2 P_2 \right) {\ln}({a^2}\,{p^2}) -
      {{{{{\ln}^2({a^2}\,{p^2})}}}\over 4} \right) \right)
\cr
\noalign{\hrule}
F_{pq}(ap)
&
\left( {1 \over {16 \pi^2}} \right)^2 \left( {a^2}\,{p^2}\,\left( 1.613 +
   \left( -{5 \over 8} + {\pi^2 \over 2} \left( P_1 - 8 P_2 \right) \right) {\ln}({a^2}\,{p^2}) +
      {{{{{\ln}^2({a^2}\,{p^2})}}}\over 8} \right) \right)
\cr
\noalign{\hrule}
F_{qk}(a p)
&
\left( {1 \over {16 \pi^2}} \right)^2 \left( -9.246 + {a^2}\,{p^2}\,
     \left( 0.9368 - {{{\ln}({a^2}\,{p^2})}\over 4} \right) \right)
\cr
\noalign{\hrule}
E_{ppkq}(ap)
&
\left( {1 \over {16 \pi^2}} \right)^2 \left( {a^2}\,{p^2}\,\left( -4.286 +
 \left({3 \over 2} - \pi^2 \left(P_1 - 8 P_2 \right) \right) {\ln}({a^2}\,{p^2}) -
      {{{{{\ln}^2({a^2}\,{p^2})}}}\over 4} \right) \right)
\cr
\noalign{\hrule}
E_{pkpq}(a p)
&
\left( {1 \over {16 \pi^2}} \right)^2 \left( {a^2}\,{p^2}\,\left( -1.490 +
  \left( {9 \over 16} - {\pi^2 \over 4} \left( P_1 - 8 P_2 \right) \right) {\ln}({a^2}\,{p^2}) -
      {{{{{\ln}^2({a^2}\,{p^2})}}}\over {16}} \right) \right)
\cr
\noalign{\hrule}
E_{pkkq}(ap)
&
\left( {1 \over {16 \pi^2}} \right)^2 \left( {a^2}\,{p^2}\,\left( 1.097 +
   \left( -{5 \over 8} + {\pi^2 \over 2} \left( P_1 - 8 P_2 \right) \right) {\ln}({a^2}\,{p^2}) +
      {{{{{\ln}^2({a^2}\,{p^2})}}}\over 8} \right) \right)
\cr
\noalign{\hrule}
G_{pppk}(ap)
&
\left( {1 \over {16 \pi^2}} \right)^2 \left( -3 \,\zeta(3)\,{a^2}\,{p^2} \right)
\cr
\noalign{\hrule}
G_{ppkq}(ap)
&
\left( {1 \over {16 \pi^2}} \right)^2 \left( {a^2}\,{p^2}\,\left( -4.214 + {\ln}({a^2}\,{p^2}) \right) \right)
\cr
\noalign{\hrule}
G_{pkpk}(ap)
&
\left( {1 \over {16 \pi^2}} \right)^2 \left( {a^2}\,{p^2}\,\left( 5.337 +
 \left( -{7 \over 8} + {\pi^2 \over 2} \left( P_1 - 8 P_2 \right) \right) {\ln}({a^2}\,{p^2}) +
      {{{{{\ln}^2({a^2}\,{p^2})}}}\over 8} \right) \right)
\cr
\noalign{\hrule}
G_{pkpq}(ap)
&
\left( {1 \over {16 \pi^2}} \right)^2 \left( {a^2}\,{p^2}\,\left( -2.732 +
      {{{\ln}({a^2}\,{p^2})}\over 4} \right) \right)
\cr
\noalign{\hrule}
G_{pkkq}(ap)
&
\left( {1 \over {16 \pi^2}} \right)^2 \left( {a^2}\,{p^2}\,\left( 2.846 +
  \left( -{{17} \over {16}} + {\pi^2 \over 4} \left( P_1 - 8 P_2 \right) \right) {\ln}({a^2}\,{p^2}) +
      {{{{{\ln}^2({a^2}\,{p^2})}}}\over {16}} \right) \right)
\cr
\noalign{\hrule}
G_{kqkq}(ap)
&
\left( {1 \over {16 \pi^2}} \right)^2 \left( 6.486 + {a^2}\,{p^2}\,
     \left( 0.1405 - {{3\,{\ln}({a^2}\,{p^2})}\over 8} \right) \right)
\cr
\noalign{\hrule}
}}


\begin{thebibliography}{99}
\bibitem{lw234} M. L\"uscher and P. Weisz, Nucl. Phys. {\bf B452}
(1995) 234.
\bibitem{lw429} M. L\"uscher and P. Weisz, Nucl. Phys. {\bf B445}
(1995) 429.
\bibitem{kawai} H. Kawai, R. Nakayama and K. Seo, Nucl. Phys.
{\bf B189} (1981) 40.
\bibitem{npb} B. All\'es, M. Campostrini, A. Feo and H. Panagopoulos,
Nucl. Phys. {\bf B413} (1994) 553.
\bibitem{plb} B. All\'es, M. Campostrini, A. Feo and H. Panagopoulos,
Phys. Lett. {\bf B324} (1994) 433.
\bibitem{tarasov} O. V. Tarasov, A. A. Vladimirov and A. Zharkov,
Phys. Lett. {\bf B93} (1980) 429.
\bibitem{abbot} L. F. Abbott, Nucl. Phys. {\bf B185} (1981) 189.
\bibitem{lw213} M. L\"uscher and P. Weisz, Nucl. Phys. {\bf B452}
(1995) 213.
\bibitem{ellism} R. K. Ellis and G. Martinelli, Nucl. Phys.
{\bf B235 [FS11]} (1984) 93.
\bibitem{caswell} W. E. Caswell and F. Wilczek, Phys. Lett. {\bf B49}
(1974) 291.
\bibitem{ellis} R. K. Ellis, Proceedings ``Gauge Theory on a
Lattice: 1984'' (1984), pg. 191.
\bibitem{parisi} G. Parisi in {\it High Energy physics-1980},
Proceedings of the XXth International AIP Conference, edited by L. Durand and
L. G. Pondrom, pg. 1531; G. Martinelli, G. Parisi and R. Petronzio, Phys. Lett.
{\bf B100} (1981) 485.
\bibitem{gcrossi} A. Di Giacomo and G. C. Rossi, Phys. Lett. {\bf B100}
(1981) 481.
\bibitem{vicari} H. Panagopoulos and E. Vicari, Nucl. Phys. {\bf B332}
(1990) 261.
\bibitem{amsterdam} B. All\'es, M. Campostrini, A. Feo and H. Panagopoulos,
Nucl. Phys. {\bf B30} (Proc. Suppl.) (1993) 243.
\end{thebibliography}
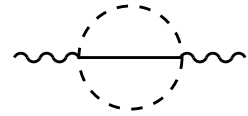
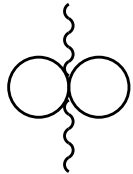
\end{document}